\documentclass[aps, prd, preprintnumbers, showpacs, nofootinbib, amsmath, amssymb, 
floatfix,
onecolumn, letterpaper, 
noshowpacs, showkeys, superscriptaddress]{revtex4}
\usepackage[utf8x]{inputenc}
\usepackage{mathtools}
\usepackage{amsfonts}
\usepackage{amsmath}
\usepackage{amssymb,epsf}
\usepackage{latexsym}
\usepackage{graphicx,epsfig}
\usepackage{amssymb}
\usepackage{subfigure}

\usepackage[citecolor=purple,linkcolor=blue,urlcolor=blue,colorlinks=true,filecolor=green,
pdfstartview=FitV, 
pdftitle={},
pdfauthor={Mojtaba Najafizadeh},
pdfsubject={},
pdfkeywords={},
pdfpagemode=None,
bookmarksopen=true]{hyperref}
\usepackage{epstopdf}
\usepackage{color}
\usepackage[normalem]{ulem}
\usepackage{slashed}

\usepackage{pdfpages}

\usepackage{eqnarray,amsmath,amssymb,mathrsfs,multirow,color}
\usepackage{amsmath}
\usepackage{amsfonts}

\usepackage{makecell} 
\usepackage{graphicx}
\usepackage{graphicx}
\graphicspath{ {images/} }
\usepackage{commath}
\usepackage{amsthm,amssymb,amsmath,mathrsfs}
\usepackage{mathtools}
\usepackage{empheq}
\usepackage{tabu} 
\usepackage[most]{tcolorbox}
\usepackage{fixmath}

\DeclarePairedDelimiterX\braket[2]{\langle}{\rangle}{#1 \delimsize\vert #2}
\graphicspath{ {images/} }

\def\be {\begin{equation}}
	\def\ee {\end{equation}}
\def\bea {\begin{eqnarray}}
	\def\eea {\end{eqnarray}}
\def\bc {\begin{center}}
	\def\ec {\end{center}}
\def\bg {\begin{align}}
	\def\eg {\end{align}}
\def\bi {\begin{itemize}}
	\def\ei {\end{itemize}}
\def\bm {\begin{pmatrix}}
	\def\em {\end{pmatrix}}

\def\le {\left}
\def\ri {\right}
\def\p {\partial}

\def\ds{\partial\hspace{-6pt}\slash\hspace{0pt}}
\def\na  {\nabla}
\def\ns{\nabla\hspace{-8pt}\slash\hspace{3pt}}
\def\Ds{D\hspace{-6pt}\slash\hspace{0pt}}
\def\ds{\partial\hspace{-6pt}\slash\hspace{1pt}}
\def\wb {\overbar\omega}

\def\ws{\omega\hspace{-7pt}\slash\hspace{3pt}}
\def\wbs{\wb\hspace{-7pt}\slash\hspace{3pt}}

\def\es{\eta\hspace{-5pt}\slash\hspace{0pt}}
\def\eb {\bar\eta}
\def\ebs{\eb\hspace{-5pt}\slash\hspace{0pt}}

\def\ep{\varepsilon}
\newcommand{\overbar}[1]{\mkern 1.5mu\overline{\mkern-1.5mu#1\mkern-1.5mu}\mkern 1.5mu}
\def\l {\ell}
\def\ll {\ell^{\,2}}

\def\Bb{\square_{_{AdS}}}
\def\Bf{\blacksquare_{_{AdS}}}
\def\1{_{_1}}
\def\2{_{_2}}

\def\c  {\cdot}

\def\g  {\gamma}
\def\d  {\delta}

\def\e  {\eta}

\def\m  {\mu}

\def\w {\omega}
\def\dw{\wb}
\def\dww{\wb^{\,2}}

\def\th {\theta}

\begin{document}                             
	
	\title{\Large{ Unconstrained Massless Higher Spin Supermultiplet in AdS$_4$ \\
			
			\vskip .5cm}} 

	
	
	
	\author{Mojtaba Najafizadeh }
	\email{mnajafizadeh@ipm.ir }
	\affiliation{School of Physics, Institute for Research in Fundamental Sciences (IPM), \\ P.O.Box 19395-5531, Tehran, Iran}

	\begin{abstract}
		We consider unconstrained formulation of the higher spin gauge theory in anti-de Sitter (AdS) spacetime, given by actions \cite{Segal:2001qq,Najafizadeh:2018cpu}, and provide on-shell supersymmetry transformations for the $\mathcal{N}=1$ unconstrained massless higher spin supermultiplet in four-dimensional AdS$_4$. Such an irreducible supermultiplet $(\mathrm{\Phi}_1, \mathrm{\Phi}_2 \,; \mathrm{\Psi}_1,\mathrm{\Psi}_2)$ contains a pair of bosonic fields, with opposite parity, which are generating functions (infinite collection of totally symmetric real tensor fields of all integer rank $s= 0, 1, \ldots, \infty$), as well as two fermionic fields, which have opposite signs of the AdS radius, that are spinorial generating functions (infinite tower of totally symmetric Majorana spinor-tensor fields of all half-integer spin $s={\scriptstyle \frac{1}{2}}, {\scriptstyle \frac{3}{2}}, \ldots, \infty$). 
	\end{abstract}
	%
	\preprint{IPM/P-2021/001}
	\vspace{-1.1cm}
	\maketitle
	\rule{\textwidth}{.4pt}
	\vspace{-1.1cm}
	{\hypersetup{linkcolor=black}
		\tableofcontents}
	\rule{\textwidth}{.4pt}

	%
	\section{Introduction}
	
	Among all unconstrained Lagrangian formulations for the higher spin gauge field theory, there exists a simple model describing massless free bosonic higher spin fields in $d$-dimensional (A)dS$_d$ spacetime, proposed by Segal in 2001 \cite{Segal:2001qq}. Later on, in 2018, this formulation was extended to massless free fermionic higher spin fields in $d$-dimensional (A)dS$_d$ spacetime \cite{Najafizadeh:2018cpu}, which we refer to both as ``Segal formulations''\,{\color{blue}\footnote{In the context of the continuous spin gauge field theory, there exist ``Segal-like formulations'' describing bosonic \cite{Schuster:2014hca} and fermionic \cite{Najafizadeh:2015uxa} continuous spin particles (CSPs) which respectively reduce to \cite{Segal:2001qq,Najafizadeh:2018cpu} in the helicity limit.}}. These formulations are given (in the metric-like approach) by the local and covariant action principles in which there are no constraints on the gauge fields and the gauge parameters, unlike Fronsdal \cite{Fronsdal:1978rb, Fronsdal:1978vb} and Fang-Fronsdal \cite{Fang:1978wz, Fang:1979hq} formulations involving some constraints on the gauge fields and parameters, in both Minkowski and AdS spacetimes. Reformulating the higher spin theory using the Segal formulation may have some advantages and simplifies some (un)solved problems, something that made us interested in extending and developing this formulation. To make clear simplicity of this formalism, let us focus on the Segal action \cite{Segal:2001qq} and present some features to compare them with the Fronsdal action \cite{Fronsdal:1978rb}:
	\begin{itemize}
		\item The Fronsdal action describes an arbitrary totally symmetric tensor field $\mathrm{\Phi}_{\m_1\ldots\m_s}(x)$ of integer rank $s$, while the Segal action describes a tower of totally symmetric tensor fields $\mathrm{\Phi}_{\m_1\ldots\m_s}(x)$ of all integer rank $s$, packed using an auxiliary vector $\e^\m$ into a single generating function $\mathrm{\Phi}(x,\e)=\sum_{s=0}^{\infty}\,\frac{1}{s!}\,\mathrm{\Phi}_{\m_1\ldots\m_s}(x)\,\e^{\m_1}\dots\e^{\m_s}$.
		
		\item At the level of equations of motion, the Fronsdal and Segal equations are equivalent and can be conveniently converted to each other \cite{Najafizadeh:2018cpu, Schuster:2013pta, Najafizadeh:2017acd}, while at the level of the action this equality has been shown in the Euclidean signature \cite{Schuster:2014hca} (the problem is still open for Lorentzian signature \cite{Bekaert:2017xin}).
		
		\item In the Segal action, there is a derivative of the Dirac delta function which at first glance it may seem complicated. However, it makes simple calculations due to the property $x\,\d(x)=0$, and its existence can be naively thought of as a constraint, since it means that dynamical fields live on a hypersurface in auxiliary space. 
		
		\item The Fronsdal action leads to the Euler-Lagrange equation which, in comparison to the spin-two case, is an Einstein-like equation which in turn reduces to a Ricci-like equation. However, the Euler-Lagrange equation of the Segal action directly leads to a Ricci-like equation, that is why the form of its action seems to be simpler than the Fronsdal one and consequently fewer calculations are needed.
	\end{itemize}
	
	As one of applications of this formalism simplifying calculations, we could find, in component formalism, supersymmetry (SUSY) transformations for the $\mathcal{N}=1$ unconstrained higher spin supermultiplet in four-dimensional Minkowski spacetime \cite{Najafizadeh:2019mun} so that the form of transformations were simple and compact. Nevertheless, the result was included the supersymmetry transformations of the Wess-Zumino supermultiplet $(\,{\scriptstyle 0}\,,\, {\scriptstyle 1/2}\,)$ as well as half-integer $(\,s\,,\, s\, {\scriptstyle +\, 1/2}\,)$ and integer $(\,{s\,\scriptstyle +\, 1/2}\,,\, s\, {\scriptstyle +\, 1}\,)$ spin supermultiplets \cite{Curtright:1979uz}. We note that in the framework of superspace formalism the off-shell $\mathcal{N}=1$, $d = 4$ higher superspin massless multiplets were studied first in \cite{Kuzenko:1993jp,Kuzenko:1993jq}, while its generalization to AdS space was given in \cite{Kuzenko:1994dm} (see also \cite{Gates:2013rka, Gates:2013ska} for component decomposition).
	
	\vspace{.2cm}
	
	As another application, the present work is devoted to study supersymmetrization of Segal formulation in four-dimensional AdS$_4$. In the unconstrained supermultiplet which we will refer to it as the so-called ``Segal supermultiplet'', we observe that the bosonic part contains two fields that have opposite parity and each one are a generating function which is an infinite tower of totally symmetric real tensor fields of all integer rank $s= 0, 1, \ldots, \infty$, while the fermionic part includes two spinorial generating function, which have opposite signs of the AdS radius, and are an infinite tower of totally symmetric Majorana spinor-tensor fields of all half-integer spin $s={\scriptstyle \frac{1}{2}}, {\scriptstyle \frac{3}{2}}, \ldots, \infty$: 
	\be 
	\hbox{$\mathcal{N}=1$ AdS$_4$ Segal supermultiplet} \qquad \Rightarrow \qquad \bigg(~\hbox{$\mathrm{\Phi}_1(x,\e)$\,~, $\mathrm{\Phi}_2(x,\e)$}~~;~~ \hbox{$\mathrm{\Psi}_1(x,\e)$\,~, $\mathrm{\Psi}_2(x,\e)$}~\bigg)\,. \label{Segal multi}
	\ee 
	In flat spacetime limit, which AdS radius goes to infinity, two Majorana fields can construct a Dirac field and therefore the supermultiplet \eqref{Segal multi} reduces to the one we obtained in \cite{Najafizadeh:2019mun}, i.e. the bosonic part becomes a complex field while the fermionic one comes to be a Dirac field.   
	
	\vspace{.2cm}
	
	We note that, in supersymmetrization of massive higher spins in flat spacetime \cite{Zinoviev:2007js} (see also its generalization to AdS \cite{Buchbinder:2019dof} and references therein), the author considered two massive fermions with opposite signs of mass terms. Motivated by this consideration, for massless higher spins in AdS space, we take into account two massless fermions with opposite signs of AdS radius.

	\vspace{.2cm}
	
	The layout of this paper is as follows. In section \ref{flat}, we will independently review the unconstrained supersymmetric higher spins in flat spacetime, which was obtained from the helicity limit ($\m=0$) of the supersymmetric continuous spin gauge theory \cite{Najafizadeh:2019mun}. Supersymmetry transformations will obtain in a rotated basis as well. In section \ref{AdS}, which includes our main results, we present supersymmetry transformations which leave invariant the supersymmetric unconstrained higher spin action in AdS$_4$. The conclusions are displayed in section \ref{conclu}. In appendices; we present our conventions in the appendix \ref{conv.}. In appendix \ref{WZ in ads}, we review the Wess-Zumino multiplet in AdS$_4$ which includes a manner that we followed to find unconstrained supersymmetry transformations in this work. In appendix \ref{closur}, we illustrate how the SUSY algebra closes on-shell in flat spacetime. Useful relations concerning supersymmetry in AdS and so on will be presented in the appendix \ref{Useful}.

	\section{Unconstrained higher spins in flat space} \label{flat}
	In this section, we review the supersymmetrization of unconstrained higher spins in flat spacetime using Segal formulation which was obtained from the helicity limit of the continuous spin gauge theory in \cite{Najafizadeh:2019mun}. However, here, we include more details such as expressing the supersymmetry action and SUSY transformations in terms of real fields. Results in a rotated fermionic system are presented as well. 
	
	\subsection{Bosonic and fermionic actions}
	
	In flat spacetime, Segal formalism can be given by the bosonic \cite{Segal:2001qq} and fermionic \cite{Najafizadeh:2018cpu} unconstrained higher spin actions (in the mostly plus signature for the metric) respectively
	\begin{align}
		S_{_{Flat}}^{\,b}&=\int d^4 x\, d^4\e ~\delta'(\e^2-1)~{\phi}^\dagger(x,\e)~ \mathrm{B}~{\phi}(x,\e)\,, ~~\,\qquad \qquad\quad
		\mathrm{B}:=\Box - \le(\e \cdot \p \ri)\le(\eb \cdot \p\ri)+ \tfrac{1}{2}\,(\e^2 -1\,) \le(\eb \cdot \p \ri)^2 \label{Segal action f}\\[5pt]
		S_{_{Flat}}^{\,f}&=\int d^4x\,d^4\e~\delta'(\e^2-1)~\overline{\psi}(x,\e)\,(\,\es-1\,)~\mathrm{F}~\psi(x,\e)\,,~~\qquad \mathrm{F}:=\ds \,-\, (\es +1\,) \,( \eb\cdot \p)\,,\label{Najafi action f}
	\end{align}
	where $\e^\m$ is a 4-dimensional auxiliary Lorentz vector localized to the unit hyperboloid of one sheet $\e^2=1$, $\g^\m$ are the 4-dimensional Dirac gamma matrices, $\d'$ is the derivative of the Dirac delta function with respect to its argument, i.e. $\d'(a)=\tfrac{d}{d a}\,\d(a)$, and   
	\begin{align}
		\eb_\m&:=\p/\p\,\e^\m \,,\qquad  \p_\m:=\p/\p x^\m \,, \qquad \Box:=\p^2\,, \\[5pt]
		\es&:=\g^\m\,\e_\m\,,~~~\, \qquad \ds:=\g^\m\,\p_\m\,, \qquad\,\, \overline{\psi}:=\psi^\dagger i\g^0\,.
	\end{align} 
	The bosonic complex field $\phi$ is unconstrained and introduces by a collection of totally symmetric complex tensor fields
	$\phi_{\m_1 \dots \m_s}(x)$ of all integer rank $s$, packed into a single generating function 
	\be 
	\phi(x,\e)=\sum_{s=0}^{\infty}\,\frac{1}{s!}~\e^{\m_1} \dots \e^{\m_s}~\phi_{\m_1 \dots \m_s}(x)\,. \label{phi}
	\ee 
	The fermionic Dirac field $\psi$ is unconstrained and introduces by a tower of totally symmetric Dirac spinor-tensor fields $\psi_{\m_1 \dots \m_s}(x)$ of all half-integer spin $s+\frac{1}{2}$\,, given by the generating function  
	\be 
	\psi(x,\e)=\sum_{s=0}^{\infty}\,\frac{1}{s!}~\e^{\m_1} \dots \e^{\m_s}~\psi_{\m_1 \dots \m_s}(x)\,, \label{psi}
	\ee 
	where the spinor index is left implicit. The bosonic action \eqref{Segal action f} is invariant under gauge transformations 
	\begin{align}
		\delta_{\xi_1} \phi (x,\e)&= \big[    \,\e \cdot \p  -  \tfrac{1}{2}\, (\e^2-1  )   (\eb \cdot \p   ) \big] \xi_1 (x,\e)\,,\label{gt b 1} \\[3pt]
		\delta_{\xi_2} \phi (x,\e)&=(\e^2 - 1 )^2 \, {\xi_2}(x,\e)\,,\label{gt b 2}
	\end{align}
	where $\xi_1, \xi_2$ are two arbitrary unconstrained complex gauge transformation parameters. The fermionic action \eqref{Najafi action f} is invariant under spinor gauge transformations
	\begin{align}
		\delta_{\zeta_1} \,\psi(x,\e)&=\big[\,\ds \,(\es +1 \,) - (\eta^2 -1 ) (\eb \cdot \p) \,\big]  {\zeta_1}(x,\e)\,, \label{zet 1}\\[3pt]
		\delta_{\zeta_2} \,\psi(x,\e)&= (\e^2-1) ( \es -  1\,) \, {\zeta_2}(x,\e)\,,\label{zet 2}
	\end{align}
	where $\zeta_1$, $\zeta_2$ are two arbitrary unconstrained spinor gauge transformation parameters.
	
	\subsection{Supersymmetry transformations}
	As one can see, the bosonic \eqref{phi} and fermionic \eqref{psi} unconstrained fields in Segal formulation include a tower of all spins, therefore equalizing bosonic and fermionic degrees of freedom in the supermultiplet does not make sense. However, for such supermultiplet we found that the number of real bosonic and fermionic fields should be equal. Indeed, the $\mathcal{N}=1$ Segal supermultiplet in 4-dimensional flat spacetime can be denoted by
	\be 
	\bigg(~~\phi(x,\e)~~,~~\psi(x,\e)~~\bigg) \label{Segal multiplet}
	\ee 
	where the complex bosonic field $\phi$ and the Dirac fermionic field $\psi$ have equal real fields. It is then convenient and straightforward to demonstrate that the unconstrained SUSY higher spin action 
	\be 
	S_{_{Flat}}^{^{\,SUSY}}=S_{_{Flat}}^{\,b}\,[\phi]+S_{_{Flat}}^{\,f}\,[\psi]\label{48}
	\ee 
	which is a sum of the bosonic \eqref{Segal action f} and fermionic \eqref{Najafi action f} unconstrained actions is invariant under the following supersymmetry transformations
	\begin{align}
		&\delta\,\phi(x,\e)\, =\,\tfrac{1}{\sqrt{2}}\,\, \bar\epsilon\,\,\big(\,1+\g^5\,\big)\,\big(\es+1\,\big)\,\psi(x,\e)\,,\qquad\qquad
		\delta\,\psi(x,\e)\, =-\,\tfrac{1}{\sqrt{2}}~\mathrm{X}~\big(\,1-\g^5\,\big)\,\epsilon~\phi(x,\e)\,,\label{susy t}
	\end{align}
	where $\epsilon$ is global supersymmetry parameter, which is a Dirac spinor, and the operator $\mathrm{X}$ defines as
	\be
	\mathrm{X}:=-\,\ds+\tfrac{1}{2}\,\big(\es-1\,\big)\big(\eb\c\p\big)\,.\label{X}
	\ee 
	We then can simply find that the commutator of supersymmetry transformations \eqref{susy t} on the bosonic and fermionic fields become respectively
	\begin{align}
		[\,\d_1\,,\,\d_2\,]\,\phi(x,\e)&=\,2\,(\bar\epsilon_2\,\ds\,\epsilon_1)\,\phi(x,\e)\,,\label{b clou} \\[5pt]
		[\,\d_1\,,\,\d_2\,]\,\psi(x,\e)&\approx\,2\,(\bar\epsilon_2\,\ds\,\epsilon_1)\,\psi(x,\e)\,+ \,{\footnotesize \hbox{G.T.}} \,,\label{susy f}
		%
	\end{align}
	where `` $\approx$ '' means that we have used the Euler-Lagrange equation of the fermionic action \eqref{Najafi action f}, i.e. 
	\be 
	\delta'(\e^2-1)(\,\es-1\,)\,\mathrm{F}\,\psi=0\,, \label{eom f}
	\ee
	and ``{\footnotesize G.T.}'' denotes a term proportional to the fermionic gauge transformation \eqref{zet 1} (see appendix \eqref{closur} for more detail). Theses together indicate that the SUSY algebra is closed on-shell up to a fermionic gauge transformation.

	\vspace{.5cm}
	
	\noindent\textbf{SUSY transformations in terms of real fields:} 
	
	\vspace{.2cm}
	
	\noindent For later purposes, let us take into account the complex bosonic field $\phi$ in terms of two real bosonic fields $\phi\1, \phi\2$ which would have opposite parity, and consider the Dirac spinor field $\psi$ in terms of two Majorana spinor fields $\psi\1, \psi\2$, i.e. 
	\be 
	\phi=\tfrac{1}{\sqrt{2}}\,(\phi\1-i\,\phi\2)\,,\qquad\qquad \psi=\tfrac{1}{\sqrt{2}}\,(\psi\1-i\,\psi\2)\,.\label{phi,psi}
	\ee 
	By this consideration, one can plug \eqref{phi,psi} into \eqref{48}, so as the SUSY higher spin action \eqref{48} converts to\,{\color{blue}\footnote{By inserting \eqref{phi,psi} into \eqref{48}, a factor of $1/2$ was appeared in \eqref{Susyflat}, stating that in comparison with \eqref{Segal action f}, \eqref{Najafi action f} we are now dealing with real fields.}}
	\begin{align} 
		S_{_{Flat}}^{^{\,SUSY}}
		&=\tfrac{1}{2}\,\int d^4 x\, d^4\e~\delta'(\e^2-1)\, 
		\Big[\,\phi\1\,\mathrm{B}\,\,\phi\1 \,+\,
		\phi\2\,\mathrm{B}\,\,\phi\2 \,+\,\overline{\psi}\1\,(\es-1)\,\mathrm{F}\,\psi\1
		\,+\,\overline{\psi}\2\,(\es-1)\,\mathrm{F}\,\psi\2\,\Big]\,, \label{Susyflat}
	\end{align}
	where the bosonic $\mathrm{B}$ and fermionic $\mathrm{F}$ operators were introduced in \eqref{Segal action f}, \eqref{Najafi action f}, and the gauge fields $\phi_i, \psi_i$ ($i=1,2$) are generating functions with a similar form as \eqref{phi}, \eqref{psi}, except that they are now real fields. It is then convenient to find that the rewritten SUSY action \eqref{Susyflat} is invariant under the following supersymmetry transformations 
	\begin{align}
		\d\,\phi\1&=\tfrac{1}{\sqrt{2}}\,\bar\ep\,\Big[\,(\es+1)\,\psi\1-i\,\g^5\,(\es+1)\,\psi\2\,\Big]\,,\qquad \qquad
		\d\,\psi\1=-\,\tfrac{1}{\sqrt{2}}~\mathrm{X} \,\le(\phi\1+i\,\g^5\,\phi\2\ri)\ep\,,\label{global susy01}\\[3pt]
		\d\,\phi\2&=\tfrac{1}{\sqrt{2}}\,\bar\ep\,\Big[\,(\es+1)\,\psi\2+i\,\g^5\,(\es+1)\,\psi\1\,\Big]\,,\qquad\qquad
		\d\,\psi\2=-\,\tfrac{1}{\sqrt{2}}~\mathrm{X} \,\le(\phi\2-i\,\g^5\,\phi\1\ri)\ep\,,\label{global susy02}
	\end{align}
	where the operator $\mathrm{X}$ was given by \eqref{X}. We note that since we are dealing here with real fields, the supersymmetry parameter $\ep$ is a Majorana spinor, unlike the previous case in which the supersymmetry parameter $\epsilon$ was a Dirac spinor. 
	
	\vspace{.2cm}
	
	It is again useful to check the closure of the SUSY algebra using supersymmetry transformations in \eqref{global susy01},\eqref{global susy02}. We will then find that the commutator of supersymmetry transformations \eqref{global susy01},\eqref{global susy02} on the bosonic $\phi_i$ and fermionic $\psi_i$ fields become
	\begin{align} 
		[\,\d_{1}\,,\,\d_{2}\,]\,\phi_i&=2\,(\bar\ep\2\,\ds\,\ep\1)\,\phi_i\,,~\,\,\,\qquad\qquad\qquad\qquad\qquad\, \qquad\qquad\qquad\qquad~~ i=1,2 \label{algeb1}\\
		[\,\d_{1}\,,\,\d_{2}\,]\,\psi_i&=2\,(\bar\ep\2\,\ds\,\ep\1)\,\psi_i~+\,\underset{j=1}{\overset{2}{\textstyle\sum}}\,\Big(\,{\footnotesize \hbox{G.T.}}(\psi_j)+{\footnotesize \hbox{E.O.M.}}(\psi_j)\,\Big)\,, ~\quad\qquad\qquad\, i=1,2  \label{algeb2}
	\end{align}
	where ${\footnotesize \hbox{G.T.}}(\psi_j)$ denotes a term proportional to gauge transformation of $\psi_j$
	\be 
	\delta \,\psi_j(x,\e)=\big[\,\ds \,(\es +1 \,) - (\eta^2 -1 ) (\eb \cdot \p) \,\big]  {\zeta}(x,\e)\,, 
	\ee 
	and ${\footnotesize \hbox{E.O.M.}}(\psi_j)$ stands for a term proportional to the equation of motion of $\psi_j$
	\be 
	\delta'(\e^2-1)(\,\es-1\,)\,\mathrm{F}\,\psi_j=0\,. \label{eom real}
	\ee
	As one can see, by applying the equation of motion for both fermionic fields $\psi_1, \psi_2$, the algebra closes up to two spinor gauge transformations. We note that the difference between two fermionic equations of motion in \eqref{eom f} and \eqref{eom real} is related to the fermionic fields. The former \eqref{eom f} includes a Dirac spinor $\psi$, while the latter \eqref{eom real} contains a Majorana spinor $\psi_j$.    
	
	\vspace{.5cm}
	
	\noindent\textbf{Matrix notation:} 
	
	\vspace{.2cm}
	
	\noindent Let us employ a matrix notation which not only makes notation pretty, but also enable us to present supersymmetry transformations in a new basis conveniently. In a matrix notation, bosonic and fermionic parts of the SUSY higher spin action \eqref{Susyflat} can be written in terms of column and row vectors
	\begin{align} 
		S_{_{Flat}}^{^{\,SUSY}}
		&=\tfrac{1}{2}\,\int d^4 x\, d^4\e~\delta'(\e^2-1)\, 
		\bigg[\,\begin{pmatrix}
			\phi\1 & \phi\2  
		\end{pmatrix}\begin{pmatrix}
			\mathrm{B} & 0 \\
			0 & \mathrm{B}  
		\end{pmatrix}\begin{pmatrix}
			\phi\1 \\ \phi\2  
		\end{pmatrix}\,+\,\begin{pmatrix}
			\overline{\psi}\1 & \overline{\psi}\2  
		\end{pmatrix}(\es-1)\begin{pmatrix}
			\mathrm{F} & 0 \\
			0 & \mathrm{F}  
		\end{pmatrix}
		\begin{pmatrix}
			\psi\1 \\ \psi\2 
		\end{pmatrix}\,\bigg]\,,\label{susy matrix action}
	\end{align}
	thus, supersymmetry transformations \eqref{global susy01},\eqref{global susy02} in terms of column vectors\,{\color{blue}\footnote{Note that, for example, a row vector for bosonic transformations becomes  
			\begin{align}
				\d \begin{pmatrix}
					\phi\1 &
					\phi\2  
				\end{pmatrix}&=-\,\tfrac{1}{\sqrt{2}}\,
				\begin{pmatrix}
					\overline{\psi}\1 &
					{ \overline{\psi}\2  }
				\end{pmatrix}(\eta\hspace{-4pt}\slash\hspace{0pt}-1)
				\begin{pmatrix}
					1 & i\,\g^5 \\
					{ -i\,\g^5} & { 1 }
				\end{pmatrix}\ep\,.
	\end{align}}} will take a compact form as the following
	\begin{align}
		\d \begin{pmatrix}
			\phi\1  \\
			\phi\2  
		\end{pmatrix}&=\tfrac{1}{\sqrt{2}}\,\bar\ep
		\begin{pmatrix}
			1 & { -i\,\g^5} \\
			i\,\g^5 & { 1 }
		\end{pmatrix}(\es+1)
		\begin{pmatrix}
			\psi\1  \\
			{\psi\2}  
		\end{pmatrix}\,,
		\qquad\qquad 
		\d\begin{pmatrix}
			\psi\1  \\
			\psi\2  
		\end{pmatrix}=-\,\tfrac{1}{\sqrt{2}}\,\mathrm{X}
		\begin{pmatrix}
			1 & i\,\g^5 \\
			-i\,\g^5 & 1 
		\end{pmatrix}\bm
		\phi\1  \\
		\phi\2  
		\em\ep\,,\label{susy matrix}
	\end{align}
	where the operators $\mathrm{B}, \mathrm{F}, \mathrm{X}$ were introduced in \eqref{Segal action f},\eqref{Najafi action f},\eqref{X} respectively. Moreover, the algebra of supersymmetry transformations \eqref{algeb1},\eqref{algeb2} in matrix notation can be written as
	\begin{align}
		\hspace{-.2cm}	[\,\d_1\,,\,\d_2\,] \begin{pmatrix}
			\phi\1  \\
			\phi\2  
		\end{pmatrix}&=
		\begin{pmatrix}
			\,\mathrm{\Delta}_{0} & 0 \\
			0 & \mathrm{\Delta}_{0}\,
		\end{pmatrix}
		\begin{pmatrix}
			\phi\1  \\
			\phi\2  
		\end{pmatrix}\,,\qquad
		[\,\d_1\,,\,\d_2\,] \begin{pmatrix}
			\psi\1  \\
			\psi\2  
		\end{pmatrix}\approx
		\begin{pmatrix}
			\,\mathrm{\Delta}_{0} & 0 \\
			0 & \mathrm{\Delta}_{0}\,
		\end{pmatrix}
		\begin{pmatrix}
			\psi\1  \\
			\psi\2  
		\end{pmatrix}\,+\,{\footnotesize \hbox{G.T.}}\,, \quad \hbox{with} ~\mathrm{\Delta}_{0}:=2(\bar\ep\2\,\ds\,\ep\1)
		\label{susy rotated}
	\end{align} 	
	which is closed on-shell ($\approx$) up to gauge transformations ({\footnotesize G.T.}).

	\vspace{.5cm}
	
	\noindent\textbf{SUSY transformations in a new basis:} 
	
	\vspace{.2cm}
	
	\noindent To close this section, and for later purposes, let us present above results in a new basis. To this end, if one rotates bosonic or fermionic column vectors, one then gets them in a new basis. For example, let us rotate fermionic column vector by an angle of $\theta$
	\be 
	\begin{pmatrix}
		\Psi\1  \\
		\Psi\2  
	\end{pmatrix}=\begin{pmatrix}
		\cos\th & -\sin\th \\
		\sin\th & ~\,\,\cos\th 
	\end{pmatrix}
	\begin{pmatrix}
		\psi\1  \\
		\psi\2  
	\end{pmatrix}\,, \label{rotated}
	\ee  
	where $\Psi\1, \Psi\2$ are fermionic higher spin fields in rotated basis. In this new basis, it is easy to see that the form of the supersymmetric action \eqref{susy matrix action}, for any $\theta$, will not change, except that we will have to substitute $\psi_i$ by $\Psi_i$ ($i=1,2$). However, supersymmetry transformations \eqref{susy matrix} in the new basis \eqref{rotated} become
	\begin{subequations}
		\label{susy rotated2}
		\begin{align}
			\d \begin{pmatrix}
				\phi\1  \\[4pt]
				\phi\2  
			\end{pmatrix}&=\tfrac{1}{\sqrt{2}}\,\bar\ep\,
			\begin{pmatrix}
				~\,\,\cos\theta+i\g^5\sin\th & \sin\th-i\g^5\cos\th \\[4pt]
				-\sin\th+i\g^5\cos\th & \cos\th+i\g^5\sin\th 
			\end{pmatrix}(\es+1)
			\begin{pmatrix}
				\Psi\1  \\[4pt]
				\Psi\2  
			\end{pmatrix}\,, \label{susy rotated22} \\[4pt]
			\d\begin{pmatrix}
				\Psi\1  \\[4pt]
				\Psi\2  
			\end{pmatrix}&=\tfrac{-\,1}{\sqrt{2}\,}\,\mathrm{X}
			\begin{pmatrix}
				\,\cos\theta+i\g^5\sin\th & -\sin\th+i\g^5\cos\th \\[4pt]
				\,\sin\th-i\g^5\cos\th & \,\,~\cos\th+i\g^5\sin\th 
			\end{pmatrix}\bm
			\phi\1  \\[4pt]
			\phi\2  
			\em\,\ep\,. \label{susy rotated33} 
		\end{align}
	\end{subequations}
	Nevertheless, one may check that the algebra of rotated supersymmetry transformations \eqref{susy rotated2}, for any $\th$, will close and would be as before \eqref{susy rotated}, upon substituting $\psi_i$ by $\Psi_i$ ($i=1,2$). Therefore, we conclude that in flat spacetime two set of non-rotated \eqref{susy matrix} and rotated \eqref{susy rotated2} supersymmetry transformations satisfy the SUSY algebra. The reason we here discussed rotated supersymmetry transformations is related to the next section. In fact, when we present supersymmetry transformations in anti-de Sitter space, we will see that its flat spacetime limit would be coincide with rotated transformations \eqref{susy rotated2} with $\th=\pi/4$.

	\subsection{Relation to the Fronsdal formalism}\label{fronsdal}
	As we discussed in the introduction, Segal formulation seems to be so simple and its results can be translated into constrained formalism, i.e. Fronsdal formalism. Indeed, one can show that the Segal multiplet \eqref{Segal multiplet} which is an irreducible multiplet can be written in terms of constrained fields, and then such an obtained multiplet would be reducible and can be decomposed into a direct sum of the Wess-Zumino multiplet $(\,{\scriptstyle 0}\,,\, {\scriptstyle \frac{1}{2}}\,)$; all half-integer spin supermultiplets $(\,s\,,\, s\, {\scriptstyle +\, \frac{1}{2}}\,)$, $s \geqslant 1$; and all integer spin supermultiplets $(\, s\, {\scriptstyle +\,\frac{1}{2}}\,,\,s\,{\scriptstyle +\,1}\,)$, $s \geqslant 0$, i.e.
	\be 
	\Big(~~\phi(x,\e)~,~\psi(x,\e)~~\Big)\quad \Longrightarrow \quad
	\Big(~~\phi(x,\w)~,~\psi(x,\w)~~\Big)\equiv
	\Big(~0~,~\tfrac{1}{2}~\Big)~~\oplus~~ \sum_{s=1}^{\infty}~\Big(~s~,~s+\tfrac{1}{2}~\Big)
	~~\oplus~~ \sum_{s=0}^{\infty}~\Big(~s+\tfrac{1}{2}~,~s+1~\Big)\,.\nonumber
	\ee  
	By making a relationship between the Segal multiplet and the Fronsdal one, we indeed demonstrated \cite{Najafizadeh:2019mun} that unconstrained supersymmetry transformations \eqref{susy t}, which have a very simple compact form will include the well-known supersymmetry transformations of chiral multiplet. Moreover, they will contain supersymmetry transformations of the half-integer spin supermultiplet $(\,s\,,\, s\, {\scriptstyle +\, 1/2}\,)$ \cite{Curtright:1979uz}
	\bea
	\d \,\phi_s(x,\w) &=&{\scriptstyle\sqrt{2}}\,\bar\varepsilon~{\psi}_s(x,\w)\,, \label{12ca}\\[3pt]
	\d \,\psi_s(x,\w) &=&-\, \tfrac{1}{\sqrt{2}}\,\Big[\,2\,\ds - \ws\,\tfrac{1}{(N+1)}\,(\dw\c\p)+\ws\,\ds\,\tfrac{1}{(N+1)}\,\wbs - \ws\,(\w\c\p)\,\tfrac{1}{2(N+2)}\,\dww\,\Big]\,\varepsilon~\phi_s(x,\w)\,, \label{34ca}
	\eea
	as well as the supersymmetry transformations of the integer spin supermultiplet {$(\, s\, {\scriptstyle +\, 1/2}\,,\,s\,{\scriptstyle +\,1}\,)$}\cite{Curtright:1979uz}
	\bea
	\hspace{-4.5cm}\,~~\d \,\phi_{s+1}(x,\w) &=&\bar\varepsilon\,\,\ws\,\tfrac{1}{\sqrt{(N+1)}}~{\psi}_s(x,\w)\,, \label{12cb}  \\[0pt]
	\hspace{-4.5cm}~\,~\d \,\psi_s(x,\w) &=&\tfrac{1}{\sqrt{(N+1)}}\,\Big[\,\ds\,\wbs- (\wb\c\p)-\tfrac{1}{2}\,(\w\c\p)\,\dww\,\Big]\,\varepsilon~\phi_{s+1}(x,\w)\,.
	\eea
	Here, we reviewed and brought our attention to the fact that in flat spacetime there exists a relationship between supersymmetry transformations of the Segal formulation and the Fronsdal one. This equality can be generally made in anti-de Sitter space where we present supersymmetry transformations in the next section, however, we will leave making this connection in current work.

	\section{Unconstrained higher spins in AdS$_4$ space} \label{AdS}
	In this section, which is the main part of the current work, we first present the bosonic and fermionic unconstrained higher spin actions in anti-de Sitter space, and then provide supersymmetry transformations.
	
	\subsection{Bosonic action}
	
	In four-dimensional AdS$_4$ spacetime, the unconstrained massless bosonic higher spin action can be given by \cite{Segal:2001qq}\,{\color{blue}\footnote{In AdS space, there was a typo in the operator $V_{11}$ of the action presented in \cite{Segal:2001qq} which is corrected in this paper and \cite{Najafizadeh:2018cpu}.}}
	\begin{align}
		S_{_{AdS}}^{\,b}=\frac{1}{2}\int d^4 x\, d^4\e~e ~{\mathrm{\Phi}}(x,\e) ~\delta'(\e^2-1)\, \Big( ~\mathbf{B} + \mathbf{B}_{\l}  ~\Big) \,{\mathrm{\Phi}}(x,\e)\,, \label{Segal action}
	\end{align}
	with
	\begin{align}
		& \mathbf{B}:= \Bb - \le(\e \cdot \nabla \ri)\le(\eb \cdot \nabla\ri)
		+ \tfrac{1}{2}\,(\e^2 -1\,) \le(\eb \cdot \nabla \ri)^2\,, \qquad\qquad \mathbf{B}_{\l}:= -\, \ll\, \le( N^2 - 2N -2 + \e^2\,\eb^{\,2} - 2\,\eb^{\,2} \, \ri) \,,\label{B,Bl}
	\end{align}
	where $\e^a$ is a 4-dimensional auxiliary Lorentz vector localized to the unit hyperboloid of one sheet $\e^2=1$, $\d'$ is the derivative of the Dirac delta function with respect to its argument, i.e. $\d'(a)=\tfrac{d}{d a}\,\d(a)$, and $e:= \hbox{det}\, e^a_\m$ where $e^a_\m$ stands for vielbein of AdS$_4$ space. The $\l:=1/R$ where $R$ is the AdS radius, $\na_a$ is the Lorentz covariant derivative, $\Bb$ is the d'Alembert operator of AdS, and $N:=\e\c\eb$ (see appendix \ref{conv.} for conventions). The gauge field $\mathrm{\Phi}$ is real and unconstrained given by the generating function
	\be 
	\mathrm{\Phi}(x,\e)=\sum_{s=0}^{\infty}\,\frac{1}{s!}~\e^{a_1} \dots \e^{a_s}~\mathrm{\Phi}_{a_1 \dots a_s}(x)\,, \label{Phi}
	\ee
	where $\mathrm{\Phi}_{a_1 \dots a_s}$ are covariant totally symmetric real tensor fields of anti-de Sitter spacetime with all integer rank $s= 0, 1, \ldots, \infty$, such that flat and curved indices are related to each other as: $\mathrm{\Phi}_{a_1 \dots a_s}(x)=e^{\m_1}_{a_1}\dots e^{\m_s}_{a_s}\,\mathrm{\Phi}_{\m_1 \dots \m_s}(x)$. The action \eqref{Segal action} is invariant under the following two gauge transformations
	\bea 
	&& \delta_{\xi_1} \mathrm{\Phi} (x,\e)\,=\left[    \,\e \cdot \nabla  -  \tfrac{1}{2}\, (\e^2-1\,)   (\eb\c\na) \,\right] \xi_1 (x,\e)\,,\label{gt ads1} \\[10pt]
	&& \delta_{\xi_2} \mathrm{\Phi} (x,\e)\,= (\,\e^2 - 1\, )^{\,2} \, \xi_2(x,\e)\,,\label{gt ads2}
	\eea 
	where there are no constraints on two gauge transformation parameters $\xi_1$ and $\xi_2$. Varying the action \eqref{Segal action} with respect to the gauge field yields the Euler-Lagrange equation
	\be 
	\delta'(\e^2-1)\,\big( ~\mathbf{B} + \mathbf{B}_{\l}  ~\big) \,{\mathrm{\Phi}}(x,\e)=0\,.
	\ee

	\subsection{Fermionic action}
	In four-dimensional AdS$_4$ spacetime, the unconstrained massless fermionic higher spin action can be given by \cite{Najafizadeh:2018cpu}
	\begin{align}
		S_{_{AdS}}^{\,f}&=\frac{1}{2}\int d^4x\,d^4\e~e~\overline{\mathrm{\Psi}}(x,\e)~\delta'(\e^2-1)\le(\es-1\,\ri)\Big( ~\mathbf{F} + \mathbf{F}_{\l}  ~\Big)\,\mathrm{\Psi}(x,\e)\,,\label{Najafi action}
	\end{align}
	with
	\begin{align}
		& \mathbf{F}:= \Ds \,-\, (\es +1\,) \,( \eb\cdot D)\,, \qquad\qquad \mathbf{F}_{\l}:= \frac{\,\l\,}{2}\, \Big(2 N +\es\,\ebs+3\,\ebs\,\Big) \,,\label{F,Fl}
	\end{align}
	where, in addition to the common relations in the bosonic case, here, $\g^a$ are the 4-dimensional Dirac gamma matrices, $D_a$ is spinorial covariant derivative (appendix \ref{conv.}) and
	\be 
	\Ds:=\g^a\,D_a\,,\quad \qquad \es:=\g^a\,\e_a\,, \quad\qquad \ebs:=\g^a\,\tfrac{\p}{\p\,\e^a}\,,\quad \qquad \overline{\mathrm{\Psi}}:=\mathrm{\Psi}^\dagger i\g^0\,.
	\ee
	The fermionic field $\mathrm{\Psi}$ is real and unconstrained defined by the generating function
	\be 
	\mathrm{\Psi}(x,\e)=\sum_{s=0}^{\infty}\,\frac{1}{s!}~\e^{a_1} \dots \e^{a_s}~\mathrm{\Psi}_{a_1 \dots a_s}(x)\,, \label{Psi}
	\ee
	where $\mathrm{\Psi}_{a_1 \dots a_s}$ are totally symmetric Majorana spinor-tensor fields of anti-de Sitter spacetime with all half-integer rank $s\scriptstyle{+1/2}$, so as flat and curved indices are related to each other via $\mathrm{\Psi}_{a_1 \dots a_s}(x)=e^{\m_1}_{a_1}\dots e^{\m_s}_{a_s}\,\mathrm{\Psi}_{\m_1 \dots \m_s}(x)$, and the spinor index is left implicit. The action \eqref{Najafi action} is invariant under the following two spinor gauge transformations
	\begin{align} 
		\delta_{\zeta_1} \mathrm{\Psi}(x,\e) &= \le[\,\Ds\,(\es+1\,)-(\e^2-1)\,(\eb\c D)+\,\frac{\,\l\,}{2}\,\le[\,2 \,\es + (\es-1\,)^2 \,\ebs-(\es-1\,)\,(2 N + 4\,)  \, \ri]\ri]\,\zeta_1(x,\e)\,, \label{gt ads3} \\[10pt]
		\delta_{\zeta_2} \mathrm{\Psi}(x,\e)&=(\e^2-1)(\es-1\,)\,\zeta_2(x,\e)\,,\label{gt ads4}
	\end{align} 
	where two spinor gauge transformation parameters $\zeta_1, \zeta_2$ are unconstrained. If one varies the action \eqref{Najafi action} with respect to the gauge field $\overline{\mathrm{\Psi}}$, we will arrive at the equation of motion 
	\be 
	\delta'(\e^2-1)\,(\es-1\,)\big( ~\mathbf{F} + \mathbf{F}_{\l}  ~\big) \,{\mathrm{\Psi}}(x,\e)=0\,.\label{eom fer}
	\ee

	\vspace{.2cm}
	
	As we discussed in the introduction, we note that the form of the bosonic and fermionic actions in Segal formulation seem simpler than Fronsdal formalism. However, by taking a look at gauge transformations \eqref{gt ads1},\eqref{gt ads2},\eqref{gt ads3},\eqref{gt ads4}, one can see that, unlike the actions, gauge transformations in Segal formalism look like more complicated than Fronsdal one.

	\subsection{Killing spinors}
	In flat spacetime, global supersymmetry transformation parameter $\epsilon$ is a constant, satisfying $\p_\m \,\epsilon=0$. In anti-de Sitter space\,{\color{blue}\footnote{For a discussion about de Sitter supersymmetry refer to  this paper \cite{Anous:2014lia} }}, the supersymmetry transformations of the fields are proportional to a spinor parameter $\ep(x)$, which is a Killing spinor in the anti-de Sitter space, i.e. $\ep(x)$ must satisfy the Killing spinor equation \cite{Burges:1985qq} (see \cite{deWit:1999ui} for a review of supersymmetry in AdS). Therefore, in the mostly plus signature for the metric, the Majorana Killing spinor equation in four dimensions can be defined in a straightforward manner
	\be 
	\Big(D_a+\frac{\l}{2}\,\g_a\Big)\ep(x)=0\,,
	\ee 
	where $D_a$ is spinorial covariant derivative, $\l$ is the inverse of AdS radius, and $\ep$ is a Majorana Killing spinor. By defining a modified covariant derivative as
	$ 
	\hat{D}_a\, \ep \equiv \big( D_a+\frac{\l}{2}\,\g_a\big)\,\ep=0\,,
	$
	one can check the integrability condition 
	$[\,\hat{D}_a\,,\,\hat{D}_b\,]\,\ep=0$\,, provided choosing the curvature in anti-de Sitter space as
	$ 
	R_{abcd}=-\,\ll\,(\e_{ac}\,\e_{bd}-\e_{ad}\,\e_{bc})\,.
	$ 
	
	\vspace{.2cm}
	Now let us discuss the case we are dealing with in this paper. As we mentioned, to formulate supersymmetric unconstrained massless higher spins in AdS, we will have to choose two Majorana fermions with opposite signs of AdS radius. For this purpose, let us consider a form of the action \eqref{Najafi action} for each Majorana fermion which are distinguished through their AdS radius. Therefore, for two fermions $\mathrm{\Psi}_1, \mathrm{\Psi}_2$ there would be respectively two Majorana Killing spinors $\ep, \chi$ satisfying the following Killing spinor equations
	\begin{align} 
		\Big(D_a+\frac{\l_1}{2}\,\g_a\Big)\ep(x)=0\,,\label{Kiling 1}\\
		\Big(D_a+\frac{\l_2}{2}\,\g_a\Big)\chi(x)=0\,, \label{Kiling 2}
	\end{align}
	where $\l_1, \l_2$ are inversed AdS radius of fermions $\mathrm{\Psi}_1, \mathrm{\Psi}_2$ respectively. If one multiplies the equation \eqref{Kiling 1} by $\g^5$ to the left
	\be 
	\Big(D_a-\frac{\l_1}{2}\,\g_a\Big)(\g^5)\,\ep(x)=0\,,
	\ee
	and compare it with \eqref{Kiling 2}, it gives us a relationship between two Killing spinors
	\be 
	\chi=\g^5\,\ep\,,\label{re}
	\ee 
	provided we choose $\l_1=-\,\l_2=\l$. Making use of this relationship which illustrates that two Killing spinors are not independent is the fact we will apply in the next subsection. We note that, in flat spacetime, where $\l_1, \l_2 \rightarrow 0$, both Killing spinors become identical to each other $\ep=\chi$. 
	
	\subsection{Supersymmetry transformations}
	
	Now we are in a position to find supersymmetry transformations. As we already discussed, to supersymmetrize unconstrained formulation of the higher spin gauge theory in 4-dimensional AdS$_4$ spacetime for the $\mathcal{N}=1$ supermultiplet, we will consider a supermultiplet $(~\mathrm{\Phi}_1\,~, \mathrm{\Phi}_2~;~ \mathrm{\Psi}_1\,~, \mathrm{\Psi}_2~)$ containing two bosonic higher spin fields $\mathrm{\Phi}_1, \mathrm{\Phi}_2$ which have opposite parity to each other, as well as two Majorana higher spin fields $\mathrm{\Psi}_1, \mathrm{\Psi}_2$ with opposite signs of AdS radius\,{\color{blue}\footnote{Note that having opposite signs of AdS radius may not be interpreted as the result of a parity transformation. In fact, if one considers such an interpretation, then it breaks in the flat spacetime limit, in which the AdS radius vanishes. Instead, it can be interpreted as two fermionic fields having opposite mass-like terms vanishing in flat spacetime. We note that, in the bosonic case, two fields have opposite parity and it holds in both flat and AdS spaces.}}. Thus, let us take into account the supersymmetric unconstrained higher spin action for such a supermultiplet in AdS$_4$ as the following
	\begin{align} 
		S_{_{AdS}}^{^{\,SUSY}}&=S_{_{AdS}}^{\,b}\,[\mathrm{\Phi}_1]+S_{_{AdS}}^{\,b}\,[\mathrm{\Phi}_2]+S_{_{AdS}}^{\,f}\,[\mathrm{\Psi}_1] + S_{_{AdS}}^{\,f}\,[\mathrm{\Psi}_2]\,,\label{50}\\[8pt]
		&=\frac{1}{2}\,\int d^4 x\, d^4\e~e ~\delta'(\e^2-1) \bigg[\,\mathrm{\Phi}_1\,(\,\mathbf{B} + \mathbf{B}_{\l}\,)\,\mathrm{\Phi}_1 +
		\mathrm{\Phi}_2\,(\,\mathbf{B} + \mathbf{B}_{\l}\,)\,\mathrm{\Phi}_2 \nonumber\\[0pt]
		&\qquad\qquad\qquad\qquad\qquad\qquad~ +\,\overline{\mathrm{\Psi}}_1\,(\es-1)\,(\,\mathbf{F} + \mathbf{F}_{\l}\,)\,\mathrm{\Psi}_1
		+\overline{\mathrm{\Psi}}_2\,(\es-1)\,(\,\mathbf{F} - \mathbf{F}_{\l}\,)\,\mathrm{\Psi}_2\,\bigg]\,, \label{Susy}
	\end{align} 
	where the bosonic and fermionic fields $\mathrm{\Phi}_i, \mathrm{\Psi}_i$ ($i=1,2$) were defined as generating functions in \eqref{Phi},\eqref{Psi}, while operators $\mathbf{B}, \mathbf{B}_\l, \mathbf{F}, \mathbf{F}_\l$ were introduced in \eqref{B,Bl},\eqref{F,Fl} respectively, and the condition $\l_1=-\,\l_2=\l$ for two fermions has been applied by flipping the sign of $\mathbf{F}_\l$ in the last term of \eqref{Susy}.

	\vspace{.2cm}
	
	We find that the above action \eqref{Susy} is invariant under the following SUSY-like transformations
	\begin{subequations}
		\label{susy11}
		\begin{align}
			\d\,\mathrm{\Phi}_1&=\tfrac{1}{\sqrt{2}}\,\bar\ep\,\,\Big[\,(\es+1)\,\mathrm{\Psi}_1-i\,\g^5\,(\es+1)\,\mathrm{\Psi}_2\,\Big]\,,\qquad \d\,\mathrm{\Psi}_1=-\,{ \tfrac{1}{\sqrt{2}}}\,\big(\,\mathbf{X}\,+\,\mathbf{X}_\l\,\big)\,\big(\,\ep\,\mathrm{\Phi}_1+i\g^5\,\chi\,\mathrm{\Phi}_2\,\big) \,,\label{susy111}\\[5pt]
			\d\,\mathrm{\Phi}_2&=\tfrac{1}{\sqrt{2}}\,\bar\chi\,\Big[\,(\es+1)\,\mathrm{\Psi}_2+i\,\g^5\,(\es+1)\,\mathrm{\Psi}_1\,\Big]\,, \qquad \d\,\mathrm{\Psi}_2=-\,{ \tfrac{1}{\sqrt{2}}}\,\big(\,\mathbf{X}\,-\,\mathbf{X}_\l\,\big)\,\big(\,\chi\,\mathrm{\Phi}_2-i\g^5\,\ep\,\mathrm{\Phi}_1\,\big)\,,\label{susy222}
		\end{align}
		where 
		\be 
		\mathbf{X}:=-\,\Ds+\tfrac{1}{2}\,(\es-1)\,(\eb\c D)\,,\qquad\qquad
		\mathbf{X}_\l:=\l\,\le(N-1+\tfrac{1}{4}\,\es\,\ebs-\tfrac{5}{4}\,\ebs\,\ri)\,, \label{X,xl}
		\ee
	\end{subequations}		 
	and $\ep$, $\chi$, which can be related to each other through \eqref{re}, are local supersymmetry transformation parameters satisfying the Killing spinor equations \eqref{Kiling 1},\eqref{Kiling 2} with $\l_1=-\,\l_2=\l$. In a matrix notation, the SUSY-like transformations \eqref{susy11} will take the following form
	\begin{align}
		\d \begin{pmatrix}
			\mathrm{\Phi}\1  \\[3pt]
			\mathrm{\Phi}\2  
		\end{pmatrix}&=\tfrac{1}{\sqrt{2}}\,
		\begin{pmatrix}
			\bar\ep & { -i\,\bar\ep\,\g^5} \\[3pt]
			i\,\bar\chi\,\g^5 & { \bar\chi }
		\end{pmatrix}(\es+1)
		\begin{pmatrix}
			\mathrm{\Psi}\1  \\[3pt]
			{\mathrm{\Psi}\2}  
		\end{pmatrix}\,,
		\quad~
		\d\begin{pmatrix}
			\mathrm{\Psi}\1  \\[3pt]
			\mathrm{\Psi}\2  
		\end{pmatrix}=-\,\tfrac{1}{\sqrt{2}}\,
		\begin{pmatrix}
			\,(\mathbf{X}+\mathbf{X}_\l)\,\ep & i(\mathbf{X}+\mathbf{X}_\l)\g^5\chi \\[4pt]
			-i(\mathbf{X}-\mathbf{X}_\l)\g^5\ep & (\mathbf{X}-\mathbf{X}_\l)\,\chi 
		\end{pmatrix}\bm
		\mathrm{\Phi}\1  \\[3pt]
		\mathrm{\Phi}\2  
		\em\,,\label{susy matrix ads1}
	\end{align}
	One can then find that the commutator of SUSY-like transformations \eqref{susy matrix ads1} (or \eqref{susy11}) on the bosonic fields would be closed. However, the commutator on the fermionic fields can not be closed, which is why we called them SUSY-like transformations. Nevertheless, if one rotates the fermionic column vector, the commutator will close on fermionic fields as well.

	\vspace{.2cm}
	
	Therefore, to obtain supersymmetry transformations, let us go to a rotated basis in which fermionic column vector \eqref{rotated} has rotated by an angle of $\pi/4$. In this basis, supersymmetric unconstrained higher spin action is 	
	\begin{align} 
		\hspace{-.2cm}	S_{_{AdS}}^{^{\,SUSY}}
		&=\tfrac{1}{2}\int d^4 x\, d^4\e\,\delta'(\e^2-1)
		\bigg[\,\begin{pmatrix}
			\mathrm{\Phi}\1 & \mathrm{\Phi}\2  
		\end{pmatrix}\begin{pmatrix}
			\mathbf{B}+\mathbf{B}_\l & 0 \\
			0 & \mathbf{B} +\mathbf{B}_\l 
		\end{pmatrix}\begin{pmatrix}
			\mathrm{\Phi}\1 \\ \mathrm{\Phi}\2  
		\end{pmatrix}\,+\,\begin{pmatrix}
			\overline{\mathbf{\Psi}}\1 & \overline{\mathbf{\Psi}}\2  
		\end{pmatrix}(\es-1)\begin{pmatrix}
			\mathbf{F} & \,\,\mathbf{F}_\l \\
			\,\,\mathbf{F}_\l & \mathbf{F}  
		\end{pmatrix}
		\begin{pmatrix}
			\mathbf{\Psi}\1 \\ \mathbf{\Psi}\2 
		\end{pmatrix}\,\bigg]\,,\label{susy matrix action ro}
	\end{align}  
	where the bold symbols $\mathbf{\Psi}_i$ denote rotated fermionic fields, and operators $\mathbf{B}, \mathbf{B}_\l, \mathbf{F}, \mathbf{F}_\l$ were introduced in \eqref{B,Bl},\eqref{F,Fl} respectively. We then find that the supersymmetric action \eqref{susy matrix action ro} would be invariant under the following supersymmetry transformations	
	\begin{subequations}  
		\label{susy matrix ads rotat}	
		\begin{tcolorbox}[ams align, colback=white!98!black]
			\d \begin{pmatrix}
				\mathrm{\Phi}\1  \\[9pt]
				\mathrm{\Phi}\2  
			\end{pmatrix}&=\tfrac{1}{2}\,
			\begin{pmatrix}
				\bar\ep\,(1+i\g^5) &~ { \bar\ep\,(1-i\g^5)} \\[9pt]
				\bar\chi\,(-1+i\g^5) &~ { \bar\chi\,(1+i\g^5) }
			\end{pmatrix}(\es+1)
			\begin{pmatrix}
				\mathbf{\Psi}\1  \\[9pt]
				{\mathbf{\Psi}\2}  
			\end{pmatrix}\,,\label{susy matrix ads rotat1}
			\\[12pt] 
			\d\begin{pmatrix}
				\mathbf{\Psi}\1  \\[9pt]
				\mathbf{\Psi}\2  
			\end{pmatrix}&=-\,\tfrac{1}{2}\,
			\begin{pmatrix}
				(\mathbf{X}+\mathbf{X}_\l)\,\ep+i(\mathbf{X}-\mathbf{X}_\l)\,\g^5\ep &~ -(\mathbf{X}-\mathbf{X}_\l)\,\chi+i(\mathbf{X}+\mathbf{X}_\l)\,\g^5\chi \\[9pt]
				(\mathbf{X}+\mathbf{X}_\l)\,\ep-i(\mathbf{X}-\mathbf{X}_\l)\,\g^5\ep &~ ~\,\,\,(\mathbf{X}-\mathbf{X}_\l)\,\chi+i(\mathbf{X}+\mathbf{X}_\l)\,\g^5\chi 
			\end{pmatrix}\bm
			\mathrm{\Phi}\1  \\[9pt]
			\mathrm{\Phi}\2  
			\em\,,\label{susy matrix ads rotat2}
		\end{tcolorbox}
	\end{subequations}
	\hspace{-.5cm}
	where operators $\mathbf{X}, \mathbf{X}_\l$ were introduced in \eqref{X,xl}. As one can see in flat spacetime ($\l\rightarrow 0$) two local supersymmetry parameters $\ep, \chi$ become global and identical to each other $\ep=\chi$, in such a way that local supersymmetry transformations \eqref{susy matrix ads rotat} turn into global rotated transformations \eqref{susy rotated2} with $\th=\pi/4$.
	
	\vspace{.2cm}
	
	To demonstrate invariance of the supersymmetric action \eqref{susy matrix action ro} under supersymmetry transformations \eqref{susy matrix ads rotat}, one can conveniently apply the following obtained identities
	\begin{align}
		(\,\mathbf{B} + \mathbf{B}_{\l}\,)\,\ep\,\mathrm{\Phi}_i&=-\,(\,\mathbf{F} + \mathbf{F}_{\l}\,)(\,\mathbf{X} + \mathbf{X}_{\l}\,)\,\ep\,\mathrm{\Phi}_i\,,\qquad (\,\mathbf{B} + \mathbf{B}_{\l}\,)\,\g^5\,\ep\,\mathrm{\Phi}_i=-\,(\,\mathbf{F} - \mathbf{F}_{\l}\,)(\,\mathbf{X} - \mathbf{X}_{\l}\,)\,\g^5\,\ep\,\mathrm{\Phi}_i\,,\label{11}\\[5pt]
		(\,\mathbf{B} + \mathbf{B}_{\l}\,)\,\chi\,\mathrm{\Phi}_i&=-\,(\,\mathbf{F} - \mathbf{F}_{\l}\,)(\,\mathbf{X} - \mathbf{X}_{\l}\,)\,\chi\,\mathrm{\Phi}_i\,,~~\quad (\,\mathbf{B} + \mathbf{B}_{\l}\,)\,\g^5\,\chi\,\mathrm{\Phi}_i=-\,(\,\mathbf{F} + \mathbf{F}_{\l}\,)(\,\mathbf{X} + \mathbf{X}_{\l}\,)\,\g^5\,\chi\,\mathrm{\Phi}_i\,,\label{44}
	\end{align}
	where the Killing spinor equations \eqref{Kiling 1},\eqref{Kiling 2} with $\l_1=-\,\l_2=\l$ have been applied on the right hand sides. For the reader's convenience, we have presented useful relations in \eqref{1}-\eqref{8}, which using them one can straightforwardly demonstrate the above identities \eqref{11},\eqref{44}.
	
	\vspace{.2cm}
	
	Finally, to show the closure of the supersymmetry algebra in AdS$_4$, let us first focus on the bosonic fields. By applying some Majorana flip relations (see e.g. \cite{Freedman:2012zz}), we find that the commutator of supersymmetry transformations \eqref{susy matrix ads rotat} on the bosonic fields $\mathrm{\Phi}_i$ becomes   
	\begin{align}
		[\,\d_1\,,\,\d_2\,]\,\mathrm{\Phi}_i&=2\,\Big[\,\bar\ep\2\,\g^a\,\ep\1\,\na_a+\tfrac{\l}{2}\,\bar\ep\2\,\g^{ab}\,\ep\1\,M_{ab}\,\Big]\,\mathrm{\Phi}_i\,,\qquad\quad i=1,2\label{algeb}
	\end{align}
	which is indeed the supersymmetry algebra in anti-de Sitter spacetime. The covariant derivative $\na_a$, the operator $M_{ab}$ and $\g^{ab}$ are given in \eqref{M},\eqref{gamma}. We note that the commutator \eqref{algeb} was written in terms of $\ep$. If one uses the relation between two Killing spinors \eqref{re} and calculate the commutator in terms of $\chi$, one finds again a relation like \eqref{algeb} in which $\ep, \bar\ep, \l$ have substituted by $\chi, \bar\chi, -\l$ respectively. This illustrates that, in terms of $\chi$, the sign of $\l$ would be flipped, as one expects. The commutator of supersymmetry transformations \eqref{susy matrix ads rotat} on the fermionic fields would be closed upon using two fermionic equations of motion \eqref{eom fer} (for $\mathrm{\Psi}_1$ and $\mathrm{\Psi}_2$), up to two spinor gauge transformations (related to $\mathrm{\Psi}_1$ and $\mathrm{\Psi}_2$) which are proportional to \eqref{gt ads3} (see the form of the closure for fermions \eqref{algeb2} in flat space case). In matrix notation, the commutator of supersymmetry transformations \eqref{susy matrix ads rotat} on the fermionic fields reads  
	\begin{align}
		[\,\d_1\,,\,\d_2\,] \begin{pmatrix}
			\mathbf{\Psi}\1  \\
			\mathbf{\Psi}\2  
		\end{pmatrix}\approx
		\begin{pmatrix}
			\,\mathrm{\Delta}_{+\l} & 0 \\
			0 & \mathrm{\Delta}_{-\l}\,
		\end{pmatrix}
		\begin{pmatrix}
			\mathbf{\Psi}\1  \\
			\mathbf{\Psi}\2  
		\end{pmatrix}\,+\,{\footnotesize \hbox{G.T.}}\,,\label{ads algebra f}
	\end{align}
	where 
	\begin{align}
		\mathrm{\Delta}_{+\l}&=2\,\Big[\,\bar\ep\2\,\g^a\,\ep\1\,\na_a\,+\,\tfrac{\l}{2}\,\bar\ep\2\,\g^{ab}\,\ep\1\,M_{ab}\,\Big]\,,\qquad\qquad
		\mathrm{\Delta}_{-\l}=2\,\Big[\,\bar\chi\2\,\g^a\,\chi\1\,\na_a\,-\,\tfrac{\l}{2}\,\bar\chi\2\,\g^{ab}\,\chi\1\,M_{ab}\,\Big]\,.
	\end{align}
	This illustrates, unlike the flat space case \eqref{susy rotated}, if one rotates the fermionic column vector, then the SUSY algebra does not hold (recall the SUSY-like transformations). This in turn comes from the fact that here two fermionic fields $\mathbf{\Psi}\1, \mathbf{\Psi}\2$ are respectively related to two Killing spinors $\ep, \chi$, having opposite sign of the AdS radius. However, by taking the flat spacetime limit ($\l\rightarrow 0$) two Killing spinors become identical to each other $\ep=\chi$, such that $(\lim_{\l\rightarrow0}\mathrm{\Delta}_{\pm\l})=\mathrm{\Delta}_0$, and thus \eqref{ads algebra f} results in \eqref{susy rotated}. To be more precise, by taking the limit $\l\rightarrow 0$, the SUSY-like transformations \eqref{susy11} will then satisfy the SUSY algebra and reduce to \eqref{susy rotated2} with $\th=0$, while the supersymmetry transformations \eqref{susy matrix ads rotat} lead to \eqref{susy rotated2} with $\th=\pi/4$.   
	
	\section{Conclusions and outlook}\label{conclu}
	
	In this work, we took into account the unconstrained formalism of the higher spin gauge field theory given by the bosonic \cite{Segal:2001qq} and fermionic \cite{Najafizadeh:2018cpu} actions in 4-dimensional flat Minkowski and anti-de Sitter spacetimes. For the reader's convenience, we first reviewed flat spacetime case independently, which was already discussed by taking a limit from the continuous spin gauge theory in \cite{Najafizadeh:2019mun}. The Segal supermultiplet was included a complex bosonic field and a Dirac spinor field given by the generating functions. Supersymmetry transformations \eqref{susy t} were found so as the supersymmetry action \eqref{48} (a sum of the complex bosonic higher spin action and the Dirac higher spin action) left invariant and the supersymmetry algebra was closed on-shell. In order to compare our results with the AdS space case, we decomposed the complex and Dirac fields and presented supersymmetry transformations \eqref{global susy01},\eqref{global susy02} in terms of two Majorana fields, and two real bosonic fields which have opposite parity. Employing a matrix notation, supersymmetry transformations were given in a rotated basis as well \eqref{susy rotated22}. 
	
	\vspace{.2cm}
	Afterwards, we considered a generalization of the supersymmetric unconstrained higher spin formalism to the anti-de Sitter space case. In this case, we took into account the bosonic \cite{Segal:2001qq} and fermionic \cite{Najafizadeh:2018cpu} higher spin actions in 4-dimensional AdS$_4$ space and realized that the supermultiplet should comprise of two bosonic real higher spin fields (with opposite parity) as well as a pair of Majorana-spinor higher spin fields which have opposite sign of the AdS radius. Having opposite radii as well as Majorana Killing spinor equations \eqref{Kiling 1}, \eqref{Kiling 2} with $\l_1=-\,\l_2=\l$ were necessary to find supersymmetry transformations. We illustrated that the supersymmetry action \eqref{Susy} (containing four actions) is invariant under SUSY-like transformations \eqref{susy matrix ads1}, but the SUSY algebra can not be closed. To close the algebra, we rotated SUSY-like transformations and obtained supersymmetry transformations \eqref{susy matrix ads rotat} leaving invariant the supersymmetry action \eqref{susy matrix action ro}. The algebra closes on-shell up to spinor gauge transformations. We demonstrated that in flat spacetime ($\l\,\rightarrow\,0$), two Killing spinors became identical and thus supersymmetry transformations \eqref{susy matrix ads rotat} will reproduce supersymmetry transformations in a rotated basis \eqref{susy rotated22}.

	\vspace{.2cm}
	
	Let us briefly discuss the possible further works related to obtained results. As we mentioned, Segal formulation is one of unconstrained higher spin formalism which is local and covariant, and gave us so simple results in context of supersymmetry. There are other unconstrained higher spin formulations in the literature (see e.g. \cite{Gates:1996xs, Francia:2007qt, Buchbinder:2007ak, Buchbinder:2008ss, Buchbinder:2007vq, Chekmenev:2019ayr} and references therein) that examining of their supersymmetry may be interesting. There is also a different formulation in which similar infinite sets of higher spin fields appear \cite{Sorokin:2017irs}. Supersymmetric higher spin models constructed in hyperspace \cite{Bandos:1999qf,Bandos:2004nn,Florakis:2014kfa,Florakis:2014aaa} describe
	infinite-dimensional higher spin supermultiplets and thus differ from the conventional higher spin supermultiplets in this work and \cite{Najafizadeh:2019mun}. Cubic interaction vertices for the $\mathcal{N}=1$ arbitrary spin massless supermultiplets were discussed in \cite{Metsaev:2019dqt, Metsaev:2019aig} and it is interesting to study such interactions using Segal formulation (see \cite{Metsaev:2020gmb, Metsaev:2021bjh} for interacting massive and massless arbitrary spin fields and $\mathcal{N}=2$ supermultiplets in 3d flat space). One may generalize Segal formulation to massive higher spins in which on-shell/off-shell supersymmetry transformations will probably take a simple form as the massless ones, discussed in this work. In this regards, we note that recently an off-shell description of massive supermultiplets was found for the first time for half-integer supermultiplets \cite{Koutrolikos:2020tel}.

	
	
	\vspace{.5cm}
	
	\noindent\textbf{\centerline{Acknowledgements:}} 
	
	\vspace{.3cm}
	
	\noindent We are grateful to Bernard de Wit, Daniel Z. Freedman, Antoine Van Proeyen and Dmitri Sorokin for useful comments and correspondence. We thank Hamid Reza Afshar and Mohammad Mahdi Sheikh-Jabbari for helpful discussions, support and encouragement. The author also acknowledges Joseph Buchbinder, Konstantinos Koutrolikos, Sergei Kuzenko and Alexander Reshetnyak for their comments. We also thank the referee for valuable comments that improved this paper.

	\appendix

	\section{Conventions} \label{conv.}
	We work in 4-dimensional AdS$_4$ spacetime and use the {\bf mostly plus} signature for the flat metric tensor $\e_{ab}$\,. We denote coordinates with $x^a$ and momenta with $p^a:=-\,i\,{\p}/{\p x_a}$\,, while we define ``auxiliary coordinates'' with $\e^a$ and ``auxiliary momenta'' with $\w^a:=i\,{\p}/{\p \e_a}$\,. The Latin (flat) indices take values: $a=0,1,2,3$\,. Derivatives with respect to $\e^a$ are defined as: 
	\be 
	\eb_\m:=\frac{\p}{\p \e^\m}\quad\quad\quad\quad\quad  \eb_a:=\frac{\p}{\p \e^a}:=e^\m_a\,\eb_\m  \quad\quad\quad\quad\quad  N:=\e\c\eb 
	\ee
	where
	\be 
	[\,\eb^a\,,\,\e^b\,]=\e^{ab}\,,\qquad [\,\e^a\,,\,\e^b\,]=0\,, \quad [\,\eb^a\,,\,\eb^b\,]=0\,.
	\ee
	
	\vspace{.2cm}
	
	The {\bf{bosonic}} covariant derivative $\nabla_a$ is given by
	\be 
	\nabla_a:=e^\m_a~\nabla_\m\,, \quad\quad\quad \nabla_\m:={\p}/{\p x^\m}+\tfrac{1}{2}~\w^{ab}_\m ~ \mathrm{M}_{ab}\,,\quad\quad\quad
	\mathrm{M}^{ab}:= \e^a\,{\eb}^b -\, \e^b\,{\eb}^a\,, \label{M}
	\ee 
	where $e^\m_a$ is inverse vielbein of AdS$_4$ space$, \nabla_\m$ stands for the Lorentz covariant derivative, $\w^{ab}_\m$ is the Lorentz connection of AdS$_4$ space, and $\mathrm{M}^{ab}$ denotes the spin operator of the Lorentz algebra, while the Greek (curved) indices take values: $\m=0,1,2,3$\,. The D'Alembert operator of AdS$_4$ space $\Bb$ is defined by
	\be 
	\Bb := \nabla^a\,\nabla_a + e^\m_a\,\w^{ab}_\m \,\nabla_b\,.
	\ee 
	Flat and curved indices of the covariant totally symmetric tensor fields of AdS$_4$ spacetime are related to each other as: $\mathrm{\Phi}_{a_1 \dots a_s}(x)=e^{\m_1}_{a_1}\dots e^{\m_s}_{a_s}\,\mathrm{\Phi}_{\m_1 \dots \m_s}(x)$.
	
	\vspace{.2cm}
	
	The {\bf{fermionic}} (spinorial) covariant derivative $D_a$ is given by
	\be 
	D_a:=e^\m_a~D_\m\,, \quad\quad D_\m:={\p}/{\p x^\m}+\,\tfrac{1}{2}~\w^{ab}_\m \, \mathbb{M}_{bc}\,,\quad\quad \mathbb{M}^{ab}:= M^{ab}+\tfrac{1}{2}\,\g^{ab}\,,\quad\quad
	\g^{ab}:=\tfrac{1}{2}( \g^a\,{\g}^b -\, \g^b\,{\g}^a)\,,\label{gamma}
	\ee
	where $\g^a$ are the 4-dimensional Dirac gamma matrices satisfying the Clifford algebra $\{\,\g^a \,,\, \g^b\, \}=2\,\e^{ab}$\,, and 
	\be
	\le(\gamma^a\ri)^\dag = \gamma^0 \gamma^a \gamma^0\,, \quad\quad\quad \g^5=i\,\g^0\g^1\g^2\g^3 \quad\quad\quad(\gamma^0 )^\dag= -\, \gamma^0\,, \quad\quad\quad (\gamma^i)^\dag= + \,\gamma^i\,, \quad   (i=1,2,3)\,.
	\ee
	%

	\section{Wess-Zumino multiplet in AdS$_4$} \label{WZ in ads}
	In this appendix, we briefly review the Wess-Zumino model in AdS$_4$ in a way that is base of what we followed to find SUSY transformations in this work, and thus may be helpful for reader. The Wess-Zumino model in anti-de Sitter space was first formulated using superspace techniques in \cite{Ivanov:1979ft}, and then studied in the framework of off-shell component formalism in \cite{Burges:1985qq}.
	
	\vspace{.2cm}
	Let us consider a free massless real scalar field $A(x)$, and a free massless Majorana field $\psi(x)$ in a general curved background given by actions{\color{blue}\footnote{Majorana spinor is real and has half as many degrees of freedom in comparison to the Dirac spinor, thus the overall factor of 1/2 compared to the Dirac action appears.}}
	\be 
	S_{A}=-\,\frac{1}{2}\int\,d^4 x \,{\sqrt{-\,g}}\,\le(\,\na^\m A~\na_\m A+\, \tfrac{R}{6}\,A^2\ri)\,,
	\qquad\quad
	S_{\psi}=-\,\frac{1}{2}\int\,d^4 x \,{\sqrt{-\,g}}\,\le(\,\overline{\psi}\,\g^\m\,D_\m\psi\,\ri)\,,
	\ee 
	where $R$ is the scalar curvature, $\na_\m:=\p_\m$ is the ``covariant derivative'', $D_\m:=\p_\m+\frac{1}{4}\,w_\m^{\,~ab}\,\g_{ab}$ is the ``spinorial covariant derivative'', and $\overline{\psi}=\psi^\dagger\,i\,\g^0$. Now let us consider the Wess-Zumino multiplet in AdS$_4$, which is the maximally symmetric solution of Einstein's equations, in which the scalar curvature is $R=-\,12\,\ll$. Due to the equality of degrees of freedom in a multiplet, one should add a free massless pseudo-scalar field $B(x)$, and thus the free action of the on-shell massless Wess-Zumino multiplet in anti-de Sitter space should have form \cite{Ivanov:1980vb}
	\be 
	S=\frac{1}{2}\int\,d^4 x \,{\sqrt{-\,g}}\,\le(\,-\,\na^\m A~\na_\m A+2\ll A^2-\,\na^\m B~\na_\m B+2\ll B^2-\overline{\psi}\,\g^\m\,D_\m\psi\,\ri)\,.
	\ee 
	To begin our method, let us rewrite the latter action, by using integration by parts and using flat indices, in the following form
	\be 
	S=\frac{1}{2}\int\,d^4 x ~ e\le[\,A\le(\,\Bb+2\,\ll\ri)A+B\le(\,\Bb+2\,\ll\ri)B-\,\overline{\psi}\,\g^a D_a \psi \, \ri]\,,\label{WZ}
	\ee 
	where $e:= \hbox{det}\, e^a_\m$ such that $e^a_\m$ stands for vielbein of AdS$_4$ space, $\Bb:=\na^\m\,\na_\m=\na^a\,\na_a +w_a^{~\,ab}\,\na_b$ while $\na_a:=\p_a$, and $D_a:=\p_a+\frac{1}{4}\,w_a^{\,~bc}\,\g_{bc}$. 
	We then vary the action \eqref{WZ} with respect to fields $A, B, \psi$, which is proportional to
	\be 
	\d S \propto \le[\,\d A\le(\,\Bb+2\,\ll\ri)A+\d B\le(\,\Bb+2\,\ll\ri)B-\,\overline{\psi}\,\g^a D_a \,\d\psi \,+\,\cdots ~ \ri]\,,\label{WZ1}
	\ee
	in which we have kept variation of bosonic fields which are on the left-hand-side, as well as variation of fermionic field which is in the right-hand-side, while other variations will appear in dotted terms. By this approach, we will see later that the remnant variations (denoted by the dotted terms in the latter) would be Hermitian conjugation of previous terms, which will appear in the final step. In this sense, it is enough to demonstrate that existing terms in \eqref{WZ1} cancel each other by choosing suitable supersymmetry transformations. For this purpose, we consider variation of bosonic fields as ones in the flat space case 
	\begin{align}
		\d A&=\overline{\psi}\,\epsilon 
		\,,\qquad\quad \d B=i\,\overline{\psi}\,\g^5\,\epsilon 
		\,, \label{var1}
	\end{align} 
	with the difference that supersymmetry parameter $\epsilon=\epsilon(x)$ is local here, and take into account variation of fermionic field as the following ansatz
	\begin{align}
		\d \psi&=X(A\,\epsilon)+i\g^5\,Y(B\,\epsilon)\,, \label{var2}
	\end{align}  
	where $X$ and $Y$ are unknown spinorial operators which we would like to find. In \eqref{WZ1}, if one wants the dotted terms appear at the end as Hermitian conjugation of their previous terms, one needs to choose a property for $X, Y$ as $X^\dagger=-\,\g^0\,X\,\g^0$ and $Y^\dagger=-\,\g^0\,Y\,\g^0$, and define the Hermitian conjugation rule as $(\p_{a})^\dagger:=-\,\p_{a}$. This guides us to consider the most general forms for $X, Y$ as
	\be 
	X=a\,\Ds +b\,\l\,,\qquad Y=c\,\Ds +d\,\l\,, \label{X,Y}
	\ee 
	where $a, b, c, d$ are real parameters that should be determined so that the variation of the action vanishes. Plugging \eqref{var1} and the ansatz \eqref{var2} (with considering $X, Y$ as \eqref{X,Y}) into \eqref{WZ1}, one arrives at
	\begin{align}
		\d S &\propto \bigg[~\overline{\psi}\,\epsilon\le(\,\Bb+2\,\ll\ri)A+\,i\,\overline{\psi}\,\g^5\,\epsilon\le(\,\Bb+2\,\ll\ri)B\nonumber\\
		&~~~\,-\,\overline{\psi}\,\le(a\,\Ds^{\,2} +b\,\l\,\Ds\,\ri)A\,\epsilon\,+\,i\,\overline{\psi}\,\g^5\,\le(c\,\Ds^{\,2} +d\,\l\,\Ds\,\ri)B\,\epsilon \,+\,\cdots ~ \bigg]\,.\label{WZ2}
	\end{align}
	In this stage, one should act the spinorial covariant derivative $D_a$ on $A\,\epsilon$ (and $B\,\epsilon$) yielding
	\be 
	D_a\,(A\,\epsilon)=(\na_a\,A)\,\epsilon+A\,(-\,\tfrac{\l}{2}\,\g_a\,\epsilon)\,,
	\ee 
	such that we have used the Killing spinor equation
	\be 
	D_a\,\epsilon=-\,\frac{\,\l\,}{2}\,\g_a\,\epsilon\,.
	\ee 
	This makes us able to compute the action of operator $\Ds$ (and $\Ds^{\,2}$) on $A\,\epsilon$ (and $B\,\epsilon$) which yields following identities
	\begin{align} 
		\Ds\,(A\,\epsilon)&=(\,\ns\,A\,)\,\epsilon-2\,\l\,A\,\epsilon\,,\\[5pt]
		\Ds^{\,2}\,(A\,\epsilon)&=(\,D^a\,D_a+w_a^{\,~ab}\,D_b+3\,\ll\,)\,(A\,\epsilon)~=~(\,\Bb\,A\,)\,\epsilon+4\,\ll\,A\,\epsilon-\l\,(\,\ns\,A\,)\,\epsilon\,.
	\end{align}
	Therefore, given these identities, the relation \eqref{WZ2} with considering of the dotted terms results in
	\begin{align}
		\d S &\propto \bigg\{~\overline{\psi}~\le[\,(1-a)(\,\Bb\,A\,)+(\,2-4a+2b\,)A\,\ll+\l(a-b)(\ns A)\,\ri]\epsilon\nonumber\\
		&~\,+i\,\overline{\psi}\,\g^5\,\le[\,(1+c)(\,\Bb\,B\,)+(\,2+4c-2d\,)B\,\ll+\l(d-c)(\ns B)\,\ri]\epsilon \,+\,h.c. ~ \bigg\}\,.\label{WZ3}
	\end{align}
	As we already mentioned, the dotted terms in \eqref{WZ1}, \eqref{WZ2} were appeared here as Hermitian conjugation ($h.c.$) of their previous terms. Finally, at a glance, one finds that the action's variation \eqref{WZ3} vanishes by setting real parameters as 
	\be 
	a=b=1\,, \qquad\qquad c=d=-\,1\,.
	\ee 
	Substituting these parameters in \eqref{X,Y}, we find operators $X, Y$, and consequently the ansatz \eqref{var2}, which together with bosonic variations \eqref{var1} 
	\begin{align}
		\d A&~=~\overline{\psi}\,\epsilon~=~\overline{\epsilon}\,\psi\,, \\[5pt]
		\d B&~=\,i\,\overline{\psi}\,\g^5\,\epsilon~=\,i\,\overline{\epsilon}\,\g^5\,\psi\,, \\[5pt]
		\d\psi&~=~\Ds\,\big[\,(A+i\,\g^5\,B)\,\epsilon\,\big]+\l\,(A-i\,\g^5\,B)\,\epsilon~=~\big[\,\ds\,(A+i\,\g^5\,B)\,\big]\,\epsilon-\l\,(A-i\,\g^5\,B)\,\epsilon\,,
	\end{align}
	are supersymmetry transformations of the Wess-Zumino action \eqref{WZ} in AdS$_4$. This strategy (i.e. keeping variation of the left-hand-side-bosons, and considering variation of the right-hand-side-fermions) is the base of what we followed in this work to find unconstrained SUSY transformations.

	\section{Commutator of SUSY transformations} \label{closur}
	The commutator of supersymmetry transformations \eqref{susy t} on the bosonic field is simple and obvious \eqref{b clou}, while the one on the fermionic field becomes 
	\begin{align}
		[\,\d_1\,,\,\d_2\,]\,\psi(x,\e)&=\,2\,(\bar\epsilon_2\,\ds\,\epsilon_1)\,\psi(x,\e)+\,\Big[\,\ds \,(\es + 1 \,) - (\eta^2 -1 ) (\eb \cdot \p) \,\Big] \Big[\,\tfrac{1}{2}\,\bar\epsilon_1\,\g_\m\,\epsilon_2\,\g^\m\,(1-\g^5\,)\,\psi(x,\e)\,\Big]\label{feom}\\
		&+\tfrac{1}{4}\,(\bar\epsilon_2\,\g_\m\,\epsilon_1)\Big[\g^\m\,\es-3\g^\m+2\e^\m+\g^\m\,\es\,\g^5+2\e^\m\,\g^5+\g^\m\,\g^5\Big]\Big[\ds-(\es+1)\,\eb\c\p\Big]\,\psi(x,\e)\,. \nonumber%
	\end{align} 
	In the first line of the latter, if one chooses a field dependent fermionic gauge transformation parameter as $$\zeta(\psi)=\tfrac{1}{2}\,\bar\epsilon_1\,\g_\m\,\epsilon_2\,\g^\m\,(1-\g^5\,)\,\psi\,,$$ then the fermionic gauge transformation \eqref{zet 1} will appear. In the second line of \eqref{feom}, the Euler-Lagrange equation of the fermionic action \eqref{Najafi action f}, i.e.
	\be
	\delta'(\e^2-1)\le(\es-1\,\ri)\le[\, \ds \,-\, (\es +1\,) \,( \eb\cdot \p)\,\ri]\psi(x,\e)=0\,,\label{eom}
	\ee
	can be easily emerged, if one multiplies the commutator \eqref{feom} by $\d(\e^2-1)$ to the left and uses the following property of the Dirac delta function
	$$\d(\e^2-1)=-\,(\es+1)\,\d'(\e^2-1)\,(\es-1)\,.$$
	We note that after applying the fermionic equation of motion \eqref{eom} the second line in \eqref{feom} vanishes, and the Dirac delta function $\d(\e^2-1)$ can be dropped from both sides of \eqref{feom}, thus the commutator will look like as the one in \eqref{susy f}. 
	
	\section{Useful relations} \label{Useful}
	
	Since the Killing spinor equation is given by 
	\be 
	D_a\,\ep=-\,\tfrac{\l}{2}\,\g_a\,\ep\,,
	\ee 
	one can write the following relations 
	\begin{align}
		\Ds\,\varepsilon(x) &=-\, 2\,\ell~\varepsilon(x)\,,  \\
		D_a\,D^a\,\varepsilon(x) &= \ell^2\,\varepsilon(x)+\frac{\ell}{2}\,w_a^{~\,ac}\,\g_c\,\varepsilon(x)\,,\\
		D_a\,D^a(A\,B)&=(D_a\,D^a\,A)B+2\,(D_a\,A)(D^a\,B)+A(D_a\,D^a\,B)\,.
	\end{align} 
	In 4-dimensional AdS$_4$, one can act the following operators (including the spinorial covariant derivative $D_a$) on $\mathrm{\Phi}\,\varepsilon$ which gives us the following useful relations:
	\begin{align} 
		\Ds\,(\mathrm{\Phi}\,\varepsilon)&=\big[\,\ns-2\,\l\,\big]\,\mathrm{\Phi}\,\varepsilon\label{1}\\
		(\eb\c D)(\mathrm{\Phi}\,\varepsilon)&=\big[\,\eb\c\na-\tfrac{\l}{2}\,\ebs\,\big]\,\mathrm{\Phi}\,\varepsilon\\
		\Bf(\mathrm{\Phi}\,\varepsilon)&=\big[\,{\Bb}+\ll-\l\,\ns\,\big]\,\mathrm{\Phi}\,\varepsilon\\
		\Ds\,\ebs\,(\mathrm{\Phi}\,\varepsilon)&=\big[\,\ns\,\ebs+\l\,\ebs+w^{abc}\,\g_a\,\g_b\,\eb_c\,\big]\,\mathrm{\Phi}\,\varepsilon\\
		(\eb\c D)^2\,(\mathrm{\Phi}\,\varepsilon)&=\big[\,(\eb\c\na)^2-\l\,\ebs\,(\eb\c\na)
		+\tfrac{\,\ll}{4}\,\eb^2\,\big]\,\mathrm{\Phi}\,\varepsilon\\
		(\e\c D)\,\ebs\,(\mathrm{\Phi}\,\varepsilon)&=\big[\,(\e\c\na)\,\ebs+\tfrac{\l}{2}\,\es\,\ebs-\l\,N-
		w^{abc}\,\g_a\,\e_b\,\eb_c\,\big]\,\mathrm{\Phi} \,\varepsilon\\
		\Ds\,(\eb\c D)(\mathrm{\Phi}\,\varepsilon)&=\big[\,\ns\,(\eb\c\na)-\tfrac{\l}{2}\,\ns\,\ebs-2\,\l\,\eb\c\na
		-\tfrac{\,\ll}{2}\,\ebs-\tfrac{\l}{2}\,w^{abc}\,\g_a\,\g_b\,\eb_c\,\big]\,\mathrm{\Phi}\,\varepsilon\\
		(\e\c D)(\eb\c D)\,(\mathrm{\Phi}\,\varepsilon)&=\big[\,(\e\c\na)(\eb\c\na)-\tfrac{\l}{2}\,\es\,(\eb\c\na)-\tfrac{\l}{2}\,(\e\c\na)\,\ebs-\tfrac{\,\ll}{4}\,\es\,\ebs+\tfrac{\,\ll}{2}\,N+\tfrac{\l}{2}\,w^{abc}\,\g_a\,\e_b\,\eb_c\,\big]\,\mathrm{\Phi}\,\varepsilon\label{8}
	\end{align}
	We note that in the right-hand-side of above relations, the appeared covariant derivative $\na_a$ just acts on the bosonic field $\mathrm{\Phi}$, not the Killing spinor $\ep$. 
	
	\vspace{.2cm}
	
	One can also show
	\begin{align}
		\ns^{\,2}&=\nabla_a\,\nabla^a+\tfrac{1}{4}\,\g^{ab}\,R_{abcd}\,M^{cd}\\
		&=\nabla_a\,\nabla^a-\tfrac{1}{2}\,\ll\,\g^{ab}\,{M}_{ab}\\
		&=\nabla_a\,\nabla^a-\ll\,(\,\es\,\ebs-N\,)
	\end{align}
	\begin{align}
		\Ds^{\,2}&=\Ds\,\Ds= D_a\,D^a+w_a^{~\,ab}\,D_b+\tfrac{1}{4}\,\g^{ab}\,R_{abcd}\,\mathbb{M}^{cd}\\
		&=\Bf-\tfrac{1}{2}\,\ll\,\g^{ab}\,\mathbb{M}_{ab}\\
		&=\Bf-\ll\,\le[\,\es\,\ebs-N-3\,\ri]
	\end{align} 
	where
	\be 
	\Bf:=D_a\,D^a+w_a^{~\,ab}\,D_b\,.
	\ee
	In addition, we have the following useful commutation relations
	\be 
	[\,\p_a\,,\,\p_b\,]=\Omega_{ab}^{~~\,c}\,\p_c
	\ee 
	\begin{align}
		[\,\na_a\,,\,\na_b\,]&=\Omega_{ab}^{~~\,c}\,\na_c + \tfrac{1}{2}\,R_{abcd}\,M^{cd}\\
		&= \Omega_{ab}^{~~\,c}\,\na_c -\ll\,M_{ab}
	\end{align}
	\begin{align}
		[\,\eb\c\na\,,\,\e\c\na\,]&=\na_a\,\na^a+w_a^{~\,ab}\,\na_b-\frac{1}{4}\,M^{ab}\,R_{abcd}\,M^{cd}\\
		&=~\Bb~-\,\ll\,\le[\,N^2+2N-\e^2\,\eb^2\,\ri]
	\end{align}

	\begin{align}
		[\,\na_a\,,\,\e^b\,]&=-\,w_a^{~\,bc}\,\e_c \,,\qquad[\,\e^2\,,\,\na_b\,]=0\,,\qquad[\,\eb^2\,,\,\e\c \na\,]=2\,\eb\c \na\,,\\[5pt]
		[\,\na_a\,,\,\eb^b\,]&=-\,w_a^{~\,bc}\,\eb_c \,,\qquad[\,\eb^2\,,\,\na_b\,]=0\,,\qquad[\,\eb\c \na\,,\,\e^2\,]=2\,\e\c \na\,. 
	\end{align}
	
	\begin{align}
		[\,D^a\,,\,\e^b\,]&=-\,w^{abc}\,\e_c\,,\qquad
		[\,D^a\,,\,\eb^b\,]=-\,w^{abc}\,\eb_c\,,\qquad
		[\,D^a\,,\,\g^b\,]=-\,w^{abc}\,\g_c\,.
	\end{align}
	\be 
	[\,\eb^2\,,\,\e\c D\,]=2\,\eb\c D\,,\qquad
	[\,\eb\c D\,,\,\e^2\,]=2\,\e\c D \,,\qquad
	[\,\ebs\,,\,\e\c D\,]=\Ds\,,\qquad
	[\,\eb\c D\,,\,\es\,]=\Ds\,.
	\ee 
	\be 
	\{\,\Ds\,,\,\es\,\}=2\,\e\c D \,,\qquad
	\{\,\Ds\,,\,\ebs\,\}=2\,\eb\c D\,.
	\ee 
	\begin{align}
		[\,D_a\,,\,D_b\,]&=\Omega_{ab}^{~~\,c}\,D_c + \tfrac{1}{2}\,R_{abcd}\,\mathbb{M}^{cd}\\
		&= \Omega_{ab}^{~~\,c}\,D_c -\ll\,\mathbb{M}_{ab}\\
		&= \Omega_{ab}^{~~\,c}\,D_c -\ll\,{M}_{ab}-\tfrac{1}{2}\,\ll\,\g_{ab}
	\end{align}
	
	\begin{align}
		[\,\eb\c D\,,\,\e\c D\,]&=D_a\,D^a+w_a^{~\,ab}\,D_b-\tfrac{1}{4}\,M^{ab}\,R_{abcd}\,\mathbb{M}^{cd}\\
		&=\Bf+\tfrac{1}{2}\,\ll\,M^{ab}\,\mathbb{M}_{ab}\\
		&=\Bf+\ll\,\le[\e^2\,\eb^2+\tfrac{1}{2}\,\es\,\ebs-N^2-\tfrac{5}{2}\,N\ri]
	\end{align}
	
	\begin{align}
		[\,\Ds\,,\,\e\c D\,]&=\tfrac{1}{2}\,\g^a\,\e^b\,R_{abcd}\,\mathbb{M}^{cd}=-\,\ll\le[\,\es\,(N+\tfrac{3}{2})-\e^2\,\ebs\,\ri]\,,
	\end{align}
	
	\begin{align}
		[\,\eb\c D\,,\,\Ds\,]&=-\,\tfrac{1}{2}\,\g^a\,\eb^b\,R_{abcd}\,\mathbb{M}^{cd}=-\,\ll\le[\,(N+\tfrac{3}{2})\,\ebs-\es\,\eb^2\,\ri]\,.
	\end{align}

\end{document}